\documentstyle[12pt]{article}
\catcode`\@=11

\@addtoreset{equation}{section}
\def\theequation{\arabic{section}.\arabic{equation}}

\catcode`\@=11

\def\section{\@startsection{section}{1}{\z@}{.5cm plus 1ex minus
   .2ex}{.35cm plus .2ex}{\large\bf}}
\def\eqnarray{\let\@currentlabel=\theequation\refstepcounter{equation}
    \global\@eqnswtrue
    \global\@eqcnt\z@\tabskip\@centering\let\\=\@eqncr
    $$\halign to \displaywidth\bgroup\@eqnsel\hskip\@centering
      $\displaystyle\tabskip\z@{##}$&\global\@eqcnt\@ne
       \hfil${{}##{}}$\hfil
      &\global\@eqcnt\tw@ $\displaystyle\tabskip\z@{##}$\hfil
       \tabskip\@centering&\llap{##}\tabskip\z@\cr}
\def\lefteqn#1{\hbox to 4\arraycolsep{$\displaystyle #1$\hss}}
\def\thesection{\arabic{section}.\hskip-.5em}

\def\appendix{\setcounter{section}{0}
        \def\thesection{Appendix.}
        \def\theequation{\Alph{section}.\arabic{equation}}}

\long\def\@makefntext#1{\parindent 0cm\noindent
\hbox to 1em{\hss$^{\@thefnmark}$}#1}
\def\IR{{\hbox{{\rm I}\kern-.2em\hbox{\rm R}}}}
\def\IH{{\hbox{{\rm I}\kern-.2em\hbox{\rm H}}}}
\def\IC{{\ \hbox{{\rm I}\kern-.6em\hbox{\bf C}}}}
\def\IZ{{\hbox{{\rm Z}\kern-.4em\hbox{\rm Z}}}}
\def\rref#1{Eq.~(\ref{#1})}
\def\H{{\cal H}}
\newcommand{\beq}{\begin{equation}}
\newcommand{\eeq}{\end{equation}}
\newcommand{\Tr}{\hbox{Tr}}
\def\ccite#1{${}^{\citen{#1}}$}
\newcommand{\NPB}[1]{{\sl Nucl.~Phys.}~{\bf B#1}}
\newcommand{\Ann}[1]{{\sl Ann.~Phys.}~{\bf #1}}
\newcommand{\CMP}[1]{{\sl Commun.~Math.~Phys.}~{\bf #1}}
\newcommand{\PLB}[1]{{\sl Phys.~Lett.}~{\bf B#1}}

\newcommand{\PRL}[1]{{\sl Phys.~Rev.~Lett.}~{\bf #1}}

\newcommand{\MPLA}[1]{{\sl Mod.~Phys.~Lett.}~{\bf A#1}}

\newcommand{\IJMPB}[1]{{\sl Int.~J.~Mod.~Phys.}~{\bf B#1}}
\newcommand{\CQG}[1]{{\sl Class.~Quant.~Grav.}~{\bf #1}}
\newcommand{\PRD}[1]{{\sl Phys.~Rev.}~{\bf D#1}}

\newcommand{\JMP}[1]{{\sl J.~Math.~Phys.}~{\bf #1}}
\begin{document}
%
%
%
%
\def\citen#1{%
\edef\@tempa{\@ignspaftercomma,#1, \@end, }
\edef\@tempa{\expandafter\@ignendcommas\@tempa\@end}%
\if@filesw \immediate \write \@auxout {\string \citation {\@tempa}}\fi
\@tempcntb\m@ne \let\@h@ld\relax \let\@citea\@empty
\@for \@citeb:=\@tempa\do {\@cmpresscites}%
\@h@ld}
%
\def\@ignspaftercomma#1, {\ifx\@end#1\@empty\else
   #1,\expandafter\@ignspaftercomma\fi}
\def\@ignendcommas,#1,\@end{#1}
%
%
\def\@cmpresscites{%
 \expandafter\let \expandafter\@B@citeB \csname b@\@citeb \endcsname
 \ifx\@B@citeB\relax 
    \@h@ld\@citea\@tempcntb\m@ne{\bf ?}%
    \@warning {Citation `\@citeb ' on page \thepage \space undefined}%
 \else
    \@tempcnta\@tempcntb \advance\@tempcnta\@ne
    \setbox\z@\hbox\bgroup 
    \ifnum\z@<0\@B@citeB \relax
       \egroup \@tempcntb\@B@citeB \relax
       \else \egroup \@tempcntb\m@ne \fi
    \ifnum\@tempcnta=\@tempcntb 
       \ifx\@h@ld\relax 
          \edef \@h@ld{\@citea\@B@citeB}%
       \else 
          \edef\@h@ld{\hbox{--}\penalty\@highpenalty \@B@citeB}%
       \fi
    \else   
       \@h@ld \@citea \@B@citeB \let\@h@ld\relax
 \fi\fi%
 \let\@citea\@citepunct
}
%
\def\@citepunct{,\penalty\@highpenalty\hskip.13em plus.1em minus.1em}%
%
%
\def\@citex[#1]#2{\@cite{\citen{#2}}{#1}}%
%
%
\def\@cite#1#2{\leavevmode\unskip
  \ifnum\lastpenalty=\z@ \penalty\@highpenalty \fi 
  \ [{\multiply\@highpenalty 3 #1
      \if@tempswa,\penalty\@highpenalty\ #2\fi 
    }]\spacefactor\@m}
\let\nocitecount\relax  
%
\vspace{.5in}
\begin{flushright}
UCD-93-15\\
NSF-ITP-93-63\\
gr-qc/9305020\\
May 1993\\
\end{flushright}
\vspace{.25in}
\begin{center}
{\large\bf
SIX WAYS TO QUANTIZE\\[.5ex]
(2+1)-DIMENSIONAL GRAVITY\footnote{Talk given at the Fifth Canadian
Conference on General Relativity and Relativistic Astrophysics, Waterloo,
Ontario, May 1993}}\\
\vspace{.15in}
{S{\sc teven} C{\sc arlip}\footnote{Permanent address:
        Department of Physics, University of California, Davis, CA
        95616}\\
       {\small\it Institute for Theoretical Physics}\\
       {\small\it University of California}\\
       {\small\it Santa Barbara, CA 93106-4030, USA}\\
       {\small\it email: carlip@dirac.ucdavis.edu}}
\end{center}
\vspace{.15cm}

\section{Introduction}

There is an old joke---feel free to adapt it to your local
politics---that if you put three socialists together in a room,
you'll end up with four political parties. The same can be said
of physicists working on quantum gravity.  Quantum gravity is hard,
and in the absence of compelling experimental or mathematical guidance,
the past forty years of research has given us a remarkable variety of
approaches to the subject.

For realistic quantum gravity in four spacetime dimensions, the best
that can be said is that not all of these approaches have been
shown to fail.  There may be cause for optimism about some lines of
research---notably loop variables and string theory---but we are still
far from having a satisfactory theory.  In this situation, a valuable
tactic is to look for simpler models that share the basic
conceptual problems of quantum gravity, while at the same time being
computationally tractable.

In the past few years, it has become increasingly clear that general
relativity in 2+1 dimensions can serve as such a
model.\ccite{DJtH,tHooft,Martinec,Wita,Witb,Carlip1}  Gravity in
2+1 dimensions has a finite number of physical degrees of freedom,
and quantum field theory is effectively reduced to quantum mechanics.
At the same time, most of the underlying problems of quantum
gravity---the need to find diffeomorphism-invariant observables, the
``problem of time,'' issues of topology and topology change, the
basic question of what it means to quantize
geometry---are still present.  And in contrast to the
(3+1)-dimensional case, in 2+1 dimensions we suffer an embarrassment
of riches: many of the existing approaches can be made to work,
but they do not all give the same answers.

The goal of this talk is to discuss six of these approaches and compare
the resulting quantum theories.  My hope is that such a comparison can
provide some insight into the problems of realistic (3+1)-dimensional
quantum gravity.  Each of these methods is worthy of a talk in its
own right, and my treatment will necessarily be cursory, but I
hope it will serve as a useful introduction.

\section{Three Approaches to Classical Gravity}

In order to investigate quantum gravity in three spacetime dimensions,
it is helpful to first understand the classical theory.  This is
particularly true because different points of view on classical gravity
suggest very different approaches to quantization.  In this section,
I will briefly summarize three techniques for solving the empty space
Einstein field equations in 2+1 dimensions.  For simplicity, I will
largely restrict my attention to the most elementary nontrivial
spacetime topology, $M = \IR\times T^2$, where the spatial topology
$T^2$ is that of a two-dimensional torus.

Let us begin with the question of why general relativity is so much
simpler in three dimensions than in four.  The easiest starting point
is a straightforward counting argument.  In four dimensions, the
spatial metric at a fixed time has six independent components,
giving six configuration space degrees of freedom per spacetime
point.  Four of these are ``gauge'' degrees of freedom, however,
which can be eliminated by a choice of four coordinates.  Two physical
degrees of freedom per point remain, corresponding roughly to the two
possible polarizations of a gravitational wave.

In three spacetime dimensions, on the other hand, the spatial metric
has three independent components, and we can choose three coordinates,
leaving no degrees of freedom per point.  This counting argument
reappears mathematically in the statement that the curvature
tensor $R_{\mu\nu\rho\sigma}$ is completely determined by the
Ricci tensor:
\beq
R_{\mu\nu\rho\sigma} = g_{\mu\rho}R_{\nu\sigma}
+ g_{\nu\sigma}R_{\mu\rho} - g_{\nu\rho}R_{\mu\sigma}
- g_{\mu\sigma}R_{\nu\rho} - {1\over2}
(g_{\mu\rho}g_{\nu\sigma} - g_{\mu\sigma}g_{\nu\rho})R .
\eeq
The field equations of general relativity thus tell us that the
curvature tensor depends algebraically on the stress-energy tensor,
and in particular vanishes in the vacuum.

At first sight, this seems to make three-dimensional gravity a
completely trivial theory, hardly likely to be a useful model.  In
a topologically trivial spacetime, this is indeed the case: the
vanishing of the curvature tensor completely determines the geometry,
and no gravitational dynamics remains.  If a spacetime contains
noncontractible curves, however, this is no longer the case.

Recall that the curvature tensor is defined by parallel transport
around closed curves.  In particular, if a curve $\gamma$ encloses
an area in which $R_{\mu\nu\rho\sigma}$ vanishes, then parallel
transport around $\gamma$ is trivial.  In a simply connected
spacetime, flatness thus guarantees that parallel transport is
path-independent, allowing a global definition of parallelism.  In a
multiply connected spacetime, on the other hand, not all closed curves
enclose areas---think of a circumference of a torus---and this
argument breaks down.  Instead, the geometry of a flat spacetime
is characterized by the results of parallel transport around such
noncontractible curves, that is, by their holonomies.

This picture suggests a natural way to construct solutions of the
(2+1)-dimensional vacuum field equations, Thurston's method of
geometric structures.\ccite{Thurston,Ep,Mess,Carlipmeas}
Any flat spacetime can be covered by contractible coordinate
patches, and in each patch, the vanishing of the curvature ensures
that the geometry is simply that of Minkowski space.  All of the
geometric information is now hidden in the transition functions
describing the overlaps between patches.  Moreover, since the metric
in each patch can be chosen to be the standard Minkowski metric,
these transition functions must be isometries of $\eta_{\mu\nu}$,
i.e., elements of the three-dimensional Poincar\'e group ISO(2,1).

Such a geometry, in which a manifold $M$ is built out of patches of
a ``model space'' $X$ glued together by isomorphisms of some structure
on $X$,  is known as a geometric structure, or $G$ structure, on
$M$.  We can rephrase the (2+1)-dimensional field equations
as a statement that empty spacetime has an ISO(2,1), or Lorentzian,
structure.  The general properties of geometric structures have been
widely studied by mathematicians, and some powerful results are
available.  In particular, it can be shown that only a small number of
transition functions are needed,\ccite{Thurston} one for each generator
of the fundamental group $\pi_1M$.  In fact, a geometric structure
determines a group homomorphism $\Gamma: \pi_1M \rightarrow G$,
where $G$ is the appropriate group of isomorphisms (for us, ISO(2,1)).
The image $\Gamma(\pi_1M)$, which tells us the transition function
around each closed curve, is known as the holonomy group of the
geometric structure.  Any two conjugate holonomy groups represent
the same geometric structure, since overall conjugation is merely
a simultaneous isometry in all of the coordinate patches.  For a
Lorentzian structure, the space of possible holonomy groups is
thus\footnote{I am glossing over a topological subtlety.  The
space of homomorphisms $\hbox{Hom}(\pi_1M,\hbox{ISO(2,1)})$
has more than one  component, and only one---that for which
the SO(2,1) projection is Fuchsian---gives rise to Lorentzian
spacetimes.\ccite{Witb,Mess,Goldman}  This is the meaning
of the subscript $0$ in \rref{strucs}.}
\begin{eqnarray}
&&{\cal M} = \hbox{Hom}_0(\pi_1M,\hbox{ISO(2,1)})/\sim\,,\nonumber\\
&&\rho_1\sim\rho_2 \quad \hbox{if}\quad \rho_2
  = h\cdot\rho_1\cdot h^{-1}\,,
  \qquad h\in \hbox{ISO(2,1)} .
\label{strucs}
\end{eqnarray}

Relativists are familiar with simple examples of such constructions.
A flat torus, for instance, can be built from Euclidean space by
gluing together the edges of a parallelogram; the holonomies are
simply the pair of translations that identify the edges.  Similarly,
the general two-dimensional version of this procedure is familiar to
string theorists: it is a form of the uniformization theorem, which
allows one to characterize a surface of genus $g>1$ by means of a
$2g$-generator subgroup of SL($2,\IR$), the isometry group of the
two-dimensional metric of constant negative curvature.

For a compact manifold, the specification of a homomorphism
$\rho\in{\cal M}$ completely determines the geometric structure, and
\rref{strucs} classifies the solutions of the vacuum Einstein
field equations.  For a more phisically realistic noncompact
topology, this is not quite the case.  But Mess has shown that
for topologies of the form $\IR\times\Sigma$ with a compact spatial
section $\Sigma$, \rref{strucs} is good enough: a homomorphism
$\rho$ determines a unique maximal domain of dependence.\ccite{Mess}

In principle, the specification of a holonomy group thus determines
the geometry of a (2+1)-dimensional spacetime.  In practice,  such a
set of data is rather difficult to work with.  In a sense, this is a
classical version of the problem of observables in quantum gravity:
the holonomies of an ISO(2,1) structure provide a complete set of
diffeomorphism-invariant observables, but it is not easy to translate
their values into more conventional information about the metric
or other familiar geometric objects.  This approach is not entirely
alien to classical general relativity, however, since it resembles
the technique of Regge calculus, in which a spacetime is built
by gluing together flat simplices.  We might therefore expect quantum
theories based on Regge calculus to have some contact with the theory
of geometric structures.

The two-dimensional analog offers one possible solution to these
problems of interpretation.  It is a standard result that if a
compact Riemann surface $\Sigma$ is characterized by a geometric
structure with holonomy group $\Gamma\subset\rm{SL}(2,\IR)$, then
it can be represented as a quotient space $\IH^2/\Gamma$, where
$\IH^2$ is the upper half-plane with its standard constant negative
curvature metric. This result gives us a constructive procedure for
determining $\Sigma$, and allows a practical computation of many of
its geometric properties.  The generalization of this result to
three-dimensional spacetimes is nontrivial, but Mess has again
provided the relevant theorems,\ccite{Mess} showing that a similar
quotient construction is possible for topologies $\IR\times\Sigma$.

For concreteness, let us apply this rather abstract argument
to the case of a spacetime with the topology $M=\IR\times T^2$.
The fundamental group of the torus---and
thus of $M$---is the abelian group $\IZ\oplus\IZ$, with one generator
for each of the two independent circumferences.  The holonomy
group must therefore by generated by two commuting Poincar\'e
transformations, say $(\Lambda_1, a_1)$ and $(\Lambda_2, a_2)$.

We first consider the SO(2,1) transformations $\Lambda_1$ and
$\Lambda_2$.  Any Lorentz transformation in 2+1 dimensions fixes a
vector $n$, and for $\Lambda_1$ and $\Lambda_2$ to commute, they
must fix the same vector.  The space of holonomies thus splits
into three topological components, according to whether $n$ is
spacelike, null, or timelike, and it is not hard to show that a
well-behaved geometric structure can occur only when $n$ is
spacelike.\ccite{Mess}  We therefore demand that either $\Lambda_1$
and $\Lambda_2$ fix a spacelike vector, say $(0,0,1)$, or
else that they both be the identity.

If $\Lambda_1 = \Lambda_2 = I$, the holonomy group is generated
by a pair of arbitrary spacelike translations $a_1$ and $a_2$, and the
spacetime is simply a static flat torus.  We can choose coordinates
such that $a_1 = (0,a,0)$ and $a_2 = (0,a\tau_1, a\tau_2)$; a
fundamental region for the holonomy group on a spatial slice is then
a parallelogram with vertices at $(0,0)$, $(a,0)$, $(a\tau_1,a\tau_2)$,
and $(a(\tau_1+1),a\tau_2)$.  The torus formed by identifying the
opposite sides of such a parallelogram is said to have modulus
$\tau = \tau_1 + i\tau_2$; its area is $a^2\tau_2$.

If instead $\Lambda_1$ and $\Lambda_2$ stabilize a spacelike vector,
both must be boosts.  We can use our remaining freedom of
overall conjugation to transform the two generators to the form
\begin{eqnarray}
H(\gamma_1): (t,x,y)&\rightarrow&(t\cosh\lambda+x\sinh\lambda,\,
           x\cosh\lambda+t\sinh\lambda,\,y+a)\nonumber\\
H(\gamma_2): (t,x,y)&\rightarrow&(t\cosh\mu+x\sinh\mu,\,
           x\cosh\mu+t\sinh\mu,\,y+b) ,
\label{transf1}
\end{eqnarray}
where $\lambda$ and $\mu$ are now unique up to the single remaining
identification $(\lambda,\mu)\sim(-\lambda,-\mu)$.  To find the desired
quotient space, it is convenient to choose new Minkowski space
coordinates $t = T^{-1}\cosh u$, $x = T^{-1}\sinh u$. The holonomies
(\ref{transf1}) then reduce to translations of $u$ and $y$, and the
quotient space is simply a torus on each surface of constant $T$.
The modulus and area of this torus are not hard to
compute:\ccite{Carlipobs}
\beq
\tau = \tau_1 + i\tau_2 =
  \left(a+{i\lambda\over T}\right)^{-1}\left(b+{i\mu\over T}\right)
\label{modhol}
\eeq
and
\beq
A(T) = {a\mu-\lambda b\over T}
\label{area}
\eeq
Furthermore, the coordinate $T$ has been been chosen so that any
surface of constant $T$ has constant mean extrinsic curvature
$\Tr K = T$.  We have thus obtained an explicit expression for the
spacetime geometry as a spatial torus with a time-dependent modulus
and area, with $\Tr K$ serving as time.

In contrast to the case of purely translational holonomies, such a
spacetime clearly has nontrivial dynamics.  It is easy to check that
\beq
\left[ \tau_1  - {1\over2} \left({\mu\over\lambda} - {b\over a}\right)
\right]^2 + \tau_2{}^2
= {1\over4} \left[{\mu\over\lambda}-{b\over a}\right]^2 ,
\label{eom}
\eeq
so the modulus describes a circular motion in the upper half plane.
This is not quite the whole story, however, since not all different
values of $\tau$ represent distinct geometries.  This is because of
the presence of ``large diffeomorphisms,'' diffeomorphisms that
cannot be smoothly deformed to the identity.  The simplest such
diffeomorphism is a Dehn twist, a transformation in which the torus
is cut open along a circumference and reglued with a $2\pi$ twist
of one edge.  The full group of large diffeomorphisms of the torus
is well understood; it is generated by a pair of transformations
$S$ and $T$, which act on the modulus as\ccite{Abikoff}
\beq
S: \tau\rightarrow-{1\over \tau}\,,\qquad T: \tau\rightarrow \tau+1,
\label{mcg1}
\eeq
or equivalently,
\begin{eqnarray}
&S&: (a,\lambda)\rightarrow(b,\mu),\quad
               (b,\mu)\rightarrow(-a,-\lambda)\\
&T&: (a,\lambda)\rightarrow(a,\lambda),\quad
               (b,\mu)\rightarrow(b+a,\mu+\lambda).
\label{mcg2}
\end{eqnarray}
These transformations are known to mathematicians as modular
transformations, or as elements of the mapping class group.  A
fundamental region for the group acting on $\tau$ is the famous
``keyhole'' region $-1/2\le\tau_1\le1/2$, $|\tau|\ge1$.

We now have a complete description of the solutions of the vacuum
Einstein equations for the topology $\IR\times T^2$.  But other
techniques may offer additional insight, and it is useful to discuss
them.  One alternative, of course, is to try to actually solve the
field equations, in either first or second order form.  The first
order form is considerably easier, since as Witten has pointed out,
the action can then be treated as a Chern-Simons action for an
ordinary gauge theory.\ccite{Wita}  In particular, if we take the
local frame ${e^a_{\ \mu}}$ and the spin connection $\omega_{a\mu}
= {1\over 2}\epsilon_{abc}\omega_{\mu}^{\ \,bc}$
as independent variables, the standard Einstein action is
\begin{eqnarray}
S &=& \int d^3x\,\epsilon^{\rho\mu\nu} e^a_{\ \rho}
        \bigl( \partial_\mu \omega_{a\nu} - \partial_\nu \omega_{a\mu}
        + \epsilon_{abc}\omega^b_{\ \mu}\omega^c_{\ \nu} \bigr)\\
  &=& 2\,\int dt \int\nolimits_\Sigma d^2x\bigl( -\epsilon^{ij}
        e_{ai}\dot\omega_{aj}
        + e^a_{\ 0} {\tilde \Theta}_a + \omega_{a0}\Theta^a \bigr)
\label{CSaction}
\end{eqnarray}
with constraints
\begin{eqnarray}
\Theta^a &=&
  {1\over 2}\epsilon^{ij}\bigl( \partial_i e_{aj} - \partial_j e_{ai}
   + \epsilon^{abc}(\omega_{bi} e_{cj} - \omega_{ci} e_{bj})\bigl) \\
{\tilde \Theta}^a &=&
  {1\over 2}\epsilon^{ij}\bigl( \partial_i \omega^a{}_j
  - \partial_j \omega^a{}_i
  + \epsilon^{abc}\omega_{bi}\omega_{cj} \bigr) .
\label{CSconstraints}
\end{eqnarray}
These constraints generate the Lie algebra ISO(2,1); moreover, as
Witten observed, the frame $e_{ai}$ and the spin connection
$\omega_{ai}$ together constitute an ISO(2,1) connection on $\Sigma$.
The conditions \hbox{${\tilde \Theta}^a = \Theta^a =0$} then force
this connection to be flat, while at the same time generating
local gauge transformations, requiring us to identify gauge-equivalent
connections.  A solution of the field equations is thus described by
an equivalence class of flat ISO(2,1) connections on $\Sigma$.

But any flat connection is determined by its holonomies---the
word is now used in its fiber bundle sense---which in turn can be
described by a homomorphism $\tilde\Gamma$ from $\pi_1\Sigma$ to
the gauge group ISO(2,1).  Moreover, gauge transformations
have the effect of conjugating $\tilde\Gamma$ by an arbitrary
group element.  We thus recover the description (\ref{strucs})
of the space of solutions, with the holonomy group $\Gamma$ of
the geometric structure now replaced by the holonomy group
$\tilde\Gamma$ of a flat ISO(2,1) connection.  This new approach
gives us a more direct relationship between the holonomy group and
more conventional geometric quantities, however: given a homomorphism
$\rho\in{\cal M}$ with image $\tilde\Gamma$, we are instructed to
find a flat connection with $\tilde\Gamma$ as its holonomy group,
perform a gauge transformation to make the translation component
$e^a{}_\mu$ nonsingular, and then interpret $(\omega,e)$ as the
spin connection and triad field of the geometry.  Unruh and
Newbury\ccite{Unruh} have recently shown how to write the action
(\ref{CSaction}) explicitly in terms of such holonomies; the result
is closely related to the Wess-Zumino-Witten action for the group
ISO(2,1).\ccite{Salomonson}

For the torus, for instance, the simplest connection with the
holonomies (\ref{transf1}) is
\beq
e^{(2)} = a\,dx + b\,dy, \qquad \omega^{(2)} = \lambda\,dx + \mu\,dy .
\eeq
The candidate triad is singular, but this can be easily fixed by an
appropriate ISO(2,1) gauge transformation; we obtain\ccite{Carlipobs}

\newlength{\abc}
\setlength{\abc}{.32\textwidth}
\begin{minipage}[t]{.5\abc} \vspace{-\abovedisplayskip}
$$\vphantom{dT\over T}$$
\end{minipage}
\begin{minipage}[t]{\abc} \vspace{-\abovedisplayskip}
\begin{eqnarray*}
e^{(0)} &=& {dT\over T^2}\,, \\
e^{(1)} &=& {1\over T}(\lambda\,dx + \mu\,dy)\,, \\
e^{(2)} &=& a\,dx + b\,dy\,,
\end{eqnarray*}
\end{minipage}
\begin{minipage}[t]{\abc} \vspace{-\abovedisplayskip}
\begin{eqnarray*}
\omega^{(0)} &=& 0\,,\vphantom{dT\over T^2} \\
\omega^{(1)} &=& 0\,,\vphantom{1\over T} \\
\omega^{(2)} &=& \lambda\,dx + \mu\,dy.
\end{eqnarray*} \vspace{\abovedisplayskip}
\end{minipage}
\begin{minipage}[t]{.44\abc} \vspace{-\abovedisplayskip}
\begin{eqnarray}
&&\vphantom{dT\over T} \nonumber\\
&&\vphantom{1\over T} \nonumber\\
&&\vphantom{e^{(2)} = a\,dx + b\,dy}
\label{triad}
\end{eqnarray}
\end{minipage}

\vspace{-4ex}
\noindent It is straightforward to show that this triad gives a spatial
metric with modulus (\ref{modhol}) and area (\ref{area}), confirming
the classical equivalence of these approaches.

Finally, we can ignore the special features of three dimensions and
try to solve the Einstein field equations in their standard metric
form.  This program is most easily carried out in ADM variables,
and has been studied carefully by Moncrief\ccite{Moncrief} and
Hosoya and Nakao.\ccite{HosNak}  The Einstein action is now
\beq
S = \int d^3x
  (-\vphantom{.}^{\scriptscriptstyle(3)}\!g)^{1/2}\>
  \vphantom{.}^{\scriptscriptstyle(3)}\!R
  = \int dt\int\nolimits_\Sigma d^2x \bigl(\pi^{ij} {\dot g}_{ij}
               - N^i\H_i -N\H\bigr),
\label{action1}
\eeq
where the momentum conjugate to $g_{ij}$ is $\pi^{ij} =
\sqrt{g}\,(K^{ij} - g^{ij}K)$, with $K^{ij}$ the extrinsic curvature
of the surface $t={\rm const.}$, and
\beq
\H_i = -2\nabla_j\pi^j_{\ i}\,, \qquad
\H = {1\over\sqrt{g}}\,g_{ij}g_{kl}(\pi^{ik}\pi^{jl}-\pi^{ij}\pi^{kl})
                 -\sqrt{g}R
\label{metricconstraints}
\eeq
are the supermomentum and super-Hamiltonian constraints.  Let us
choose York's time slicing,\ccite{York} $K = \pi/\sqrt{g} = T$,
foliating the spacetime by surfaces of constant $\Tr K$.  Moncrief
shows that such a foliation always exists for solutions of the
field equations.  By the two-dimensional uniformization theorem,
we can write the spatial metric as
\beq
g_{ij} = e^{2\lambda}\tilde g_{ij},
\eeq
where $\tilde g$ is a metric of constant curvature $k$ on $\Sigma$;
$k=1$ for the sphere, $0$ for the torus, and $-1$ for surfaces of
genus greater than one.  The conjugate momenta have a corresponding
decomposition into a trace $\pi$ and a traceless part $\tilde\pi^{ij}$,
and the Hamiltonian constraint reduces to an elliptic differential
equation for $\lambda$,
\beq
\Delta_{\tilde g}\lambda - {1\over4} T^2e^{2\lambda}
+ {1\over2}\left[ {\tilde g}^{-1}\tilde g_{ij}\tilde g_{kl}
\tilde\pi^{ik}\tilde\pi^{jl}\right]e^{-2\lambda} - {k\over2} = 0.
\label{scale}
\eeq
This equation has no solution when $k=1$, and a unique solution
otherwise.  For the torus, in particular, one finds that
\beq
e^{4\lambda} = {2\over T^2} {\tilde g}^{-1}\tilde g_{ij}\tilde g_{kl}
\tilde\pi^{ik}\tilde\pi^{jl},
\label{torussol}
\eeq
eliminating $\lambda$ as an independent degree of freedom.

As always, the flat spatial metric $\tilde g_{ij}$ is determined by
a complex modulus $\tau= \tau_1 + i\tau_2$, and we can define its
conjugate momentum by
\beq
p^\alpha = \int_\Sigma d^2x\, e^{2\lambda}\left[{\tilde\pi}^{ij}
{\partial\ \over\partial\tau_\alpha}\tilde g_{ij}\right].
\eeq
Inserting \rref{torussol} back into \rref{action1}, we obtain
a reduced phase space action
\beq
S = \int dT \left[ p^\alpha {d\tau_\alpha\over dT} - H(p,\tau)\right]
\label{ADMaction}
\eeq
with an effective Hamiltonian
\beq
H(p,\tau) = {1\over T} \Bigl[ \tau_2^{\ 2}
{\left( (p^1)^2 + (p^2)^2 \right)} \Bigr]^{1/2}
\label{Ham1}
\eeq
that describes the remaining physical degrees of freedom.

The expression in brackets in \rref{Ham1} is the square of the momentum
with respect to the constant negative curvature (Poincar\'e) metric
\beq
d\sigma^2 = \tau_2{}^{-2}d\tau d\bar\tau
\label{Poincare}
\eeq
on the torus moduli space.  Except for the square root and the rather
trival time dependence, (\ref{Ham1}) is the Hamiltonian for a free
particle moving in a curved background described by this metric.
One can easily verify that the classical solutions are simply the
geodesics in such a background, i.e., semicircles centered on the real
axis,\ccite{HosNak} precisely as expected from \rref{eom}.  The ADM
approach thus gives the same set of solutions we already obtained,
and confirms the identification (\ref{modhol}) between moduli and
holonomies.  Moreover, using \rref{modhol} we can write the
Hamiltonian as
\beq
H = {a\mu-\lambda b\over T},
\label{Hamb}
\eeq
which we recognize as the area of the surface of constant $\Tr K = T$.

\section{Quantum Theory I: Reduced ADM Phase Space Quantization}

Having understood the classical solutions for our model, we can
finally turn to the problem of quantization.  The most obvious
approach is suggested by the reduced phase space action
(\ref{ADMaction}), the
classical action for the physical degrees of freedom in the York time
slicing.\ccite{HosNak2,Carlipobs}  The number of degrees of freedom
is now finite---for the torus,
two positions and two canonical momenta---and we are left with a
straightforward problem of quantum mechanics.  We can now
simply follow our noses: we make $p=p^1+ip^2$ and $\tau$ into
operators, with the standard commutation relations
\beq
[\hat\tau_\alpha,\hat p^\beta] = i\delta^\beta_\alpha,
\eeq
acting on the Hilbert space of square integrable functions of $\tau$.
The Schr\"odinger equation for this system can be read off from
the action; it is
\beq
i{\partial \psi\over\partial T} = T^{-1}\Delta_0^{1/2}\psi,
\label{Schrod1}
\eeq
where
\beq
\Delta_0 = -\tau_2{}^2\left(
{\partial^2\ \over\partial\tau_1{}^2}
+ {\partial^2\ \over\partial\tau_2{}^2}
\right)
\label{Lap}
\eeq
is the scalar Laplacian on the upper half plane with respect to the
Poincar\'e metric.  $\Delta_0$ has no negative eigenvalues, and the
(positive) square root can be defined by a spectral decomposition;
our solutions automatically have positive ``energy.''  We have
seen that the Hamiltonian on a surface of fixed $T$ is essentially
the area, so this positive frequency condition is a restriction to
wave functions describing an expanding universe.  Such a picture
is classically consistent, since no exact solution in 2+1 dimensions
recollapses, but the generalization to 3+1 dimensions is problematic.

The one real subtlety in this formulation comes from the question of
how to treat the large diffeomorphisms (\ref{mcg1}).  Quantum
gravity should presumably be invariant under the entire group of
diffeomorphisms, and it may be argued that we should therefore
restrict ourselves to wave functions that are invariant under these
transformations.  This point of view is not undisputed---one can
argue instead that the only true invariances of the quantum theory
are those generated by the constraints---but there seems to be good
evidence that the classical limit of point particle scattering
can be recovered only if one requires full mapping class group
invariance.\ccite{Carlip1,Carlipmeas}  Modular invariant
eigenfunctions of the Laplacian (\ref{Lap}) are called weight
zero Maass forms, and have been studied extensively by mathematicians;
the exact functional forms are not known, but lowest nonzero
eigenvalues have been investigated numerically.\ccite{Maassforms}

There appears to be nothing inconsistent with this approach to
quantizing (2+1)-dimensional gravity.  The dynamics is straightforward,
and the physical significance of the fundamental operators $\hat\tau$
and $\hat p$ is relatively clear.  The method depends heavily on a
classical choice of time slicing, however, in our case the choice
of a foliation by surfaces of constant $\Tr K$.  As a result, certain
apparently reasonable questions---for instance, questions about the
geometry of different time slices---seem difficult or impossible to
ask.\ccite{Kuchar}  It is not even known whether different choices
of time slicing lead to equivalent theories.

\section{Quantum Theory II: Chern-Simons Theory/Connection
Representation}

A second straightforward approach to quantizing our system starts
with the first order Chern-Simons form of the action.\ccite{Wita,Witb}
While this technique does not have any exact analog in 3+1 dimensions,
it is closely related to Ashtekar's connection
dynamics.\ccite{Ash1,Ash2}  We now begin with the action
(\ref{CSaction}), which is already in the first order form
$p\dot q + \dots$, and impose the equal time commutation relations
\beq
[\omega_{ai}(x),e^b{}_j(x')]
= - {i\over2}\delta^b_a\epsilon_{ij}\delta^2(x-x').
\label{commute}
\eeq
It is natural to consider the $\omega_{ai}$ as ``positions'' and the
$e^b{}_j$ as ``momenta'' in this representation (hence the name
``connection dynamics''), in part because the constraint $\Theta^a$
of \rref{CSconstraints} is the ``derivative'' of the constraint
$\tilde\Theta^a$: if we perform an infinitesimal variation of
$\omega$ in $\tilde\Theta^a$, the resulting equation for
$\delta\omega$ will be precisely the constraint $\Theta^a$
evaluated at $e=\delta\omega$.  The space of solutions $(\omega,e)$
of the constraints is thus the cotangent bundle of the space of flat
spin connections $\omega$.  We shall see in the next section that
this is a consequence of the structure of the group ISO(2,1).

With this choice of variables, the $\tilde\Theta^a$ constraint tells
us that the SO(2,1) connection $\omega$ must be flat, while
$\Theta^a$ generates SO(2,1) gauge transformations, requiring
us to identify gauge-equivalent connections.  Wave functions are thus
functions of equivalence classes of flat SO(2,1) connections.  For the
torus topology, in particular, the SO(2,1) piece of the connection
is parametrized by the holonomies $\lambda$ and $\mu$, and wave
functions are thus square integrable functions $\chi(\lambda,\mu)$.
Again, we should presumably demand modular invariance, in the form of
invariance under the transformations (\ref{mcg2}).  The remaining
holonomies $a$ and $b$ now become operators, whose commutators can
be determined, for instance, by comparing \rref{triad} and
\rref{commute}; we find that
\beq
[\hat a, \hat\mu] = [\hat\lambda,\hat b] = {i\over2},
\label{commutator}
\eeq
with all other commutators vanishing.

In this first order formalism, the notion of time has rather
mysteriously disappeared.  The Hamiltonian of a Chern-Simons theory
is identically zero, and there seems to be no dynamics---we have
obtained what is commonly known as a ``frozen time''
formalism.\ccite{Kuchar}  This should not be surprising, however,
since we have already encountered the same  ``problem of time''
classically.  Our wave functions $\chi$ depend only on the holonomies
$\lambda$ and $\mu$ of a geometric structure, which describe the entire
spacetime, not a particular time slice; it is already nontrivial to
extract dynamical information at the classical level.

Classically, though, we know how to solve this problem.  A geometric
structure determines an entire spacetime, but once we have constructed
that spacetime, we can simply choose our favorite time slicing and look
at the corresponding dynamics.  {}\rref{modhol}, for instance, gives
the moduli of a surface of constant $\Tr K = T$ in terms of the
corresponding holonomies.  It is natural to carry this equation over
into the quantum theory, defining a one-parameter family of operators
\beq
\hat\tau(T) = \left(\hat a+{i\hat\lambda\over T}\right)^{-1}
\left(\hat b+{i\hat\mu\over T}\right).
\label{modop}
\eeq
The Hamiltonian (\ref{Hamb}) becomes
\beq
\hat H = {\hat a\hat\mu-\hat\lambda \hat b\over T},
\label{Hamc}
\eeq
and it is not hard to check that the operators $\hat a$, $\hat b$,
$\hat\lambda$, and $\hat\mu$ obey the correct Heisenberg equations of
motion.\footnote{I am neglecting subtleties involving operator
ordering; suffice it to say that compatibility of the
modular transformations (\ref{mcg1}) and (\ref{mcg2}) places strong
restrictions on possible
orderings.\ccite{Carlipobs,Carlipdirac,Carlipord}}

{}From this point of view, Chern-Simons quantization should be
understood as a Heisenberg picture.  Each choice of time slicing
will determine a family of ``time''-dependent operators analogous
to the moduli (\ref{modop}); the corresponding Hamiltonian, in
turn, provides us with a quantum description of dynamics in that
time slicing.  Operators such as (\ref{modop}) are manifestly
diffeomorphism-invariant, and are examples of what Rovelli calls
``evolving constants of motion.''\ccite{Rovelli}  Kucha{\v r} has
raised the important question of whether the operators coming from
different time slicings can be simultaneously made self-adjoint; this
issue is not yet resolved.\ccite{Kuchar}

It should be stressed that explicit
constructions such as (\ref{modop}) depend on our ability to solve the
field equations in the corresponding time slicing.  In 3+1 dimensions,
this will no longer be possible, and we will presumably have to develop
a suitable perturbation theory for ``time''-dependent operators.

It is natural to ask how the two quantum theories discussed so far fit
together.  In particular, we might hope that the reduced phase space
ADM theory is a Schr\"odinger picture corresponding to the connection
representation Heisenberg picture.  In other words, we can ask whether
the wave functions $\psi(T)$ of the last section are simply the
eigenfunctions of the operators $\hat\tau(T)$ of this section.

As it turns out, they are not quite.  For this simple model, it is
possible to explicitly diagonalize the operators $\hat\tau$ and
$\hat\tau^\dagger$.  The resulting wave functions satisfy a
Schr\"odinger equation of the form\ccite{Carlipdirac}
\beq
i{\partial \chi\over\partial T}
= T^{-1}\Delta_{1\over2}{}^{1/2}\chi,
\eeq
where
\beq
\Delta_{1\over2}
= \Delta_0 + i\tau_2{\partial\ \over\partial\tau_1} - {1\over4}
\eeq
is the Laplacian for Maass forms of weight one-half, essentially
one-component spinors on moduli space.  If the ADM  Schr\"odinger
equation is an ordinary square root of a Klein-Gordon equation on
moduli space, the connection representation Schr\"odinger equation
is a kind of Dirac square root.

Let me mention in passing that this analysis again involves
subtleties concerning operator ordering.  By rather unnatural
changes of ordering, we can actually obtain Laplacians acting on
Maass forms of arbitrary weight, i.e., arbitrary tensors on moduli
space.  Such changes implicitly involve new inner products, however,
and therefore alter the definition of the adjoint; in particular,
we can reproduce the ADM results only at the cost of losing
hermiticity of the holonomy operators.\ccite{Carlipord}

Finally, we must confront a set of problems arising from the
topology of the space of connections.  So far, I have only discussed
connections whose holonomies correspond to geometric structures---for
the torus, holonomies that fix a spacelike vector.  In the Chern-Simons
formulation, there seems to be no compelling reason to omit the
``nongeometrical'' connections, whose presence could be taken as a
sign of additional phases of the theory.

If we insist on mapping class group invariance, however, these extra
sectors will not appear.  Consider, for instance, the
``timelike sector'' for our $\IR\times T^2$ topology.  The hyperbolic
sines and cosines in \rref{transf1} will be replaced by ordinary
sines and cosines, and
the variables analogous to $\lambda$ and $\mu$ will be angles,
say $\theta$ and $\phi$.  Modular transformations parallel to
(\ref{mcg2}) then tell us to identify, for instance, $(\theta,\phi)$
and $(\theta,\phi + n\theta)$.  But for generic values of $\theta$,
$\phi + n\theta$ can be made arbitrarily close to any other angle.
This is a symptom of the fact that a typical orbit of the mapping
class group is dense, which in turn means that the only continuous
invariant wave functions $\psi(\theta,\phi)$ in the ``timelike sector''
are constants.

\section{Quantum Theory III: Covariant Canonical Quantization}

The methods of the previous section are closely related to a somewhat
different approach known as covariant canonical quantization.  This
approach grew out of the traditional ``covariant vs.\ canonical''
debate in quantum gravity.
We understand quantum theories most clearly in a canonical
framework.  But ordinary canonical quantization requires us to make a
choice of time from the start, violating at least the spirit of general
covariance.  On the other hand, the only standard covariant approach
we have available, the path integral background field method,
is fundamentally perturbative, a feature widely believed to be
disastrous for quantum gravity.

Covariant canonical quantization attempts to avoid this dilemma by
finding a covariant setting in which to do canonical quantization.
That setting is the space of solutions of the field
equations.\ccite{Ash3,CrnWit}  For a theory with a well-posed
initial value problem, it is easy to see that this space of
solutions is isomorphic to the phase space---any point in phase space
determines a set of initial values that can be used to construct a
solution, and any solution determines a set of initial data on an
(arbitrary but fixed) initial hypersurface.

To apply this program to (2+1)-dimensional gravity, let us return to
the description (\ref{strucs}) of the space of classical solutions.
To define
a symplectic structure on this space---a necessary first step for
quantization---we observe that the group manifold of ISO(2,1) is
already a cotangent bundle, with base space SO(2,1).  Indeed, given
two curves $\Lambda_1(t)$ and $\Lambda_2(t)$ through a common point
$\Lambda(0)$ of the SO(2,1) manifold, the ``product cotangent vector''
is
\begin{eqnarray}
\lefteqn{
\left({d\ \over dt}(\Lambda_1\Lambda_2)(t)\right)
(\Lambda_1\Lambda_2)^{-1}(t)\Biggm|_{t=0}}\\
&&=
\left({d\ \over dt}\Lambda_1(t)\right)\Lambda_1{}^{-1}(t)\Biggm|_{t=0}
+ \Lambda_1(0)\left[
\left({d\ \over dt}\Lambda_2(t)\right)\Lambda_2{}^{-1}(t)\Biggm|_{t=0}
\right]\Lambda_1{}^{-1}(0) ,\nonumber
\end{eqnarray}
which can be recognized as the usual semidirect product structure for
ISO(2,1).  It is not hard to see that the space of solutions $\cal M$
of \rref{strucs} is therefore itself a cotangent bundle (ignoring
difficult questions of smoothness), whose base space
\beq
{\cal N} = \hbox{Hom}_0(\pi_1M,\hbox{SO(2,1)})/\sim
\eeq
is the space of equivalence classes of SO(2,1) holonomies.

Such a cotangent bundle automatically carries a symplectic structure,
with points in the base space serving as ``positions'' and cotangent
vectors as ``momenta.''  For our simple model, we can again quantize
by follow our noses, replacing the corresponding Poisson brackets by
commutators.  For the torus, in particular, the base space is
parametrized by the SO(2,1) holonomies $\lambda$ and $\mu$, and
the Poisson brackets of the cotangent bundle structure lead to
commutators identical to
those of \rref{commutator}.  The analysis of the previous section
therefore goes through unchanged, but now in a framework that can be
generalized to higher dimensions.

\section{Quantum Theory IV: The Loop Representation}

In 3+1 dimensional gravity, recent work on Ashtekar's variables
and the connection representation has led to an interesting
``dual'' approach to quantization, known as the loop
representation.\ccite{Ash1,Ash2,Smolin}
This method starts with the observation that a connection on a
manifold $M$ with gauge group $G$ is essentially a map from closed
loops on $M$ into $G$.  Given any loop $\gamma$ based at a point
$p\in M$, the connection determines an element of $G$ describing the
result of parallel transport around $\gamma$.  Conversely, any set of
parallel transport matrices obeying certain algebraic constraints
determines a connection.  Connections and loops are thus dual to each
other, and a wave function $\Psi[A]$ of a connection determines a dual
wave function
\beq
\tilde\Psi[\gamma] = \int[dA]\,\Psi[A]{\cal T}^0[\gamma][A] ,
\label{looptransf}
\eeq
where
\beq
{\cal T}^0[\gamma][A] = \Tr\, P\exp\left\{\int_\gamma A \right\}
\eeq
is the holonomy of $\gamma$ (the trace removes any dependence on the
base point).

In 3+1 dimensions, the functional integral in \rref{looptransf}
requires some careful definition, and it is more useful to look
instead for representations of the algebra of the loop operators
${\cal T}^{(0)}$ and their cotangent vectors
${\cal T}^{(1)}$.\ccite{Ash1}  In 2+1 dimensions, on the
other hand, the picture is simpler; the relevant connections
are now the flat SO(2,1) connections $\omega$, and instead of loops we
need only consider homotopy classes of loops.  The loop transformation
is then well-defined, and can be investigated in detail.

For the $\IR\times T^2$ topology, any homotopy class $[\gamma]$
is labeled by two winding numbers $m$ and $n$, and loop representation
wave functions are functions $\tilde\psi(m,n)$.  It is not hard to
check that\footnote{These expressions are slightly different from
those of Ref.~\citen{Marolf} because I am taking SO(2,1) rather than
SU(1,1) traces.}
\begin{eqnarray}
{\cal T}^0[(m,n)](\lambda,\mu) &=&  1+2\cosh(m\lambda+n\mu) \nonumber\\
{\cal T}^1[(m,n)](\lambda,\mu,a,b) &=&
  2\sinh(m\lambda+n\mu)\,(m\hat a + n\hat b) ,
\end{eqnarray}
and that the standard loop operator commutators give back
\rref{commutator}.  Moreover, using the symmetry $\chi(\lambda,\mu) =
\chi(-\lambda,-\mu)$, we can recognize \rref{looptransf} as an ordinary
two-sided Laplace transform from coordinates $(\lambda,\mu)$ to
$(m,n)$,
\beq
\tilde\chi(m,n)
= 2\int\!d\lambda\!\int\!d\mu\, e^{2(m\lambda+n\mu)}\chi(\lambda,\mu) ,
\label{loopt}
\eeq
up to an overall constant that disappears if one uses SU(1,1) rather
than SO(2,1) traces.

It is now natural to ask whether the transformation (\ref{loopt}) is an
isomorphism, that is, whether the connection and loop representations
are equivalent.  As Marolf has recently shown, it is not.\ccite{Marolf}
The basic problem is that the Laplace transform (\ref{loopt}) is only
evaluated at integral values of $m$ and $n$, and information at
these points is not enough to invert the transformation.  In fact,
the loop transform has a kernel that is dense in the connection
representation Hilbert space---{\em any} wave function
$\chi(\lambda,\mu)$ is the limit of a sequence of wave functions that
each have a transform $\tilde\chi = 0$.

There is a technical way to avoid this problem, essentially by
defining the transform (\ref{loopt}) only on a subset of connection
representation wave functions, but the procedure seems rather
contrived.  Alternatively, one can define loop representation
wave functions on the ``generalized loops'' of Di Bartolo et
al.\ccite{diBart}  A generalized loop is a distribution on
$\Sigma$ that behaves roughly like a loop with fractional winding
number.  If we allow such distributions, the integers $m$ and $n$
can be replaced by continuous ``winding numbers,''
and we recover a quantization equivalent to that of the connection
representation.  But again, there seems to be no strong justification
for such a choice, and the physical significance of generalized loops
is far from clear.

One possible loophole remains.  We have not yet imposed the
condition of modular invariance, which for loop representation wave
functions requires that
\beq
\tilde\chi(m,n) = \tilde\chi(n,-m) = \tilde\chi(m-n,n) .
\eeq
I do not know whether the transformation between such loop states and
the corresponding connection states is an isomorphism; this question
deserves further study.

\section{Quantum Theory V: The Wheeler-DeWitt Equation}

Yet another common approach to quantum gravity is that of the
Wheeler-DeWitt equation.  The methods discussed so far have required
us to solve the supermomentum and super-Hamiltonian constraints
before quantizing; that is, we have quantized only the physical
degrees of freedom of the classical theory.  Following Dirac, we
could instead allow wave functions to be arbitrary functionals on
the full configuration space, and impose the constraints as operator
equations to define physical wave functions.

In the first order formalism of section 4, this procedure leads to
nothing new, essentially because the constraints are first order in
the canonical momenta.  If we start with a wave function
$\Phi[\omega]$, the constraint $\tilde\Theta^a$ of \rref{CSconstraints}
acts multiplicatively, again requiring $\Phi[\omega]$ to have its
support on the flat spin connections.  $\Theta^a$ depends on $e$ as
well as $\omega$, and we must substitute
\beq
e^b{}_j = -{i\over2} \epsilon_{jk}{\delta\ \over\delta\omega_{bk}} .
\eeq
But the resulting functional differential equation is first order,
and can be integrated exactly; it tells us merely that the wave
function must be invariant under SO(2,1) transformations of $\omega$,
reproducing our previous results.

In the metric formalism of section 3, on the other hand, the
super-Hamiltonian constraint (\ref{metricconstraints}) is second
order in the momenta, and the results are rather different.  For the
$\IR\times T^2$ topology, our initial wave functions will now be
functionals $\Psi[\lambda,\tau]$ of the scale factor $\lambda$ and
the modulus $\tau$.  The supermomentum constraints are again first
order, and merely require that $\Psi$ be invariant under spatial
diffeomorphisms.  But the ${\cal H} = 0$ constraint
must now be imposed as a functional differential equation, the
Wheeler-DeWitt equation, which for this topology takes the form
\beq
\left\{
{1\over8}e^{2\lambda}{\delta\ \over\delta\lambda}
e^{-2\lambda}{\delta\ \over\delta\lambda}
+ {1\over2}\Delta_0 + 2 e^{2\lambda}(\Delta_{\tilde g}\lambda)\right\}
  \Psi[\lambda,\tau] = 0 ,
\label{WdW}
\eeq
where $\Delta_{\tilde g}$ is the Laplacian on $\Sigma$ with respect to
the flat metric $\tilde g$, and $\Delta_0$ is again the Laplacian on
the torus moduli space.

At first sight, it is hard to see how the solutions of the
Wheeler-DeWitt equation relate to the wave functions $\psi(\tau,T)$
of section 3.  Wave functions satisfying \rref{WdW} are functions
of a ``many-fingered time'' $\lambda$, and it is not easy to extract
a single parameter $T$ to describe their evolution.  Part of the
problem is now hidden in the choice of inner product on the space of
solutions of the Wheeler-DeWitt equation.  {}\rref{WdW} is a functional
Klein-Gordon equation, whose inner product should presumably be
something like a Klein-Gordon inner product,\ccite{DeWittb} say
\beq
\langle\Psi_1\mid\Psi_2\rangle = {1\over2i} \prod_{x\in\Sigma}
\int {d^2\tau\over\tau_2{}^2}\, \left(\Psi_1^*[\lambda,\tau]
{\stackrel{\leftrightarrow}{\delta}\ \over\delta\lambda}
\Psi_2[\lambda,\tau]\right)\Biggl|_{\lambda = \lambda_0} .
\label{KGprod}
\eeq
Woodard has argued that such an inner product should be understood
as a consequence of gauge-fixing.\ccite{Woodard}
According to this interpretation, we should start with a standard
inner product
\beq
\langle\Psi_1\mid\Psi_2\rangle
= \int[d\lambda]\int{d^2\tau\over\tau_2{}^2}\,
\Psi_1^*[\lambda,\tau]\Psi_2[\lambda,\tau] .
\eeq
Since we have not fixed a time slicing, this functional integral is
divergent, and must be gauge-fixed in the standard Faddeev-Popov
manner.  If we choose a gauge $\lambda=\lambda_0$, we recover
\rref{KGprod}, complete with the functional derivative, which
arises as a Faddeev-Popov determinant.  If we choose instead a
constant mean curvature gauge $\pi/\sqrt{g} = T$, it
can be shown that the inner product takes the form
\beq
\langle\Psi_1\mid\Psi_2\rangle = \int {d^2\tau\over\tau_2{}^2}\,
\tilde\Psi_1^*(T,\tau)\tilde\Psi_2(T,\tau) ,
\eeq
where
\beq
\tilde\Psi(T,\tau) = \int [d\lambda]\,
\exp\left\{iT\int_\Sigma e^{2\lambda}\right\}\,
\nu[\lambda,\tau]\Psi[\lambda,\tau]
\label{WdWstate}
\eeq
and $\nu[\lambda,\tau]$ is a measure factor coming from the
Faddeev-Popov determinant,
\beq
\nu[\lambda,\tau] = {\det}^{1/2}\left|
\Delta_{\tilde g} - 2\Delta_{\tilde g}\lambda - T^2e^{2\lambda}\right| .
\eeq

The gauge-fixed inner product is thus determined by a set of wave
functions that depend only in $T$ and $\tau$.  The obvious question
is whether they are the same functions we found in section 3.  While
this problem has not been completely analyzed, the answer is probably
that they are not.  If we ignore the Faddeev-Popov factor
$\nu[\lambda,\tau]$ in \rref{WdWstate}, and insert \rref{WdW}, we
find that
\begin{eqnarray}
\Biggl(T^2{\partial^2\ \over\partial T^2}&+&\Delta_0\Biggr)
 \tilde\Psi(T,\tau)\\
&=& \int[d\lambda]\,\exp\left\{iT\int_\Sigma e^{2\lambda}\right\}\,
\left\{T^2\left(e^{4\lambda}
- \left(\int_\Sigma e^{2\lambda}\right)^2\right)
- 4 e^{2\lambda}\Delta_{\tilde g}\lambda \right\}\Psi[\lambda,\tau]
\nonumber.
\label{WdWeqn}
\end{eqnarray}
If the functional integral were limited to configurations with
spatially constant $\lambda$, the right-hand side of \rref{WdWeqn}
would vanish, and we would recover the square of the Schr\"odinger
equation (\ref{Schrod1}) of reduced phase space quantization.
This would
be the case if $\lambda$ were required to satisfy the classical
constraint (\ref{scale}).  Instead, however, all values of $\lambda$
contribute to the functional integral, and there is no reason to
expect the right-hand side of \rref{WdWeqn} to disappear.  Based on
preliminary calculations, it seems unlikely that the inclusion of the
measure $\nu[\lambda,\tau]$ will change this
conclusion.\ccite{CarlipWdW}
For small values of $T$, however---recall that these correspond to
late values of time---it seems that the factor $\nu[\lambda,\tau]$
strongly damps any contributions with $\Delta_{\tilde g}\lambda\ne 0$,
so the Schr\"odinger equation (\ref{Schrod1}) may still be a good
approximation.

\section{Quantum Theory VI: Lattice Approaches}

Let me conclude with a brief mention of some of the lattice
and combinatorial approaches to quantum gravity in 2+1 dimensions.
This is a topic worthy of a talk of its own, and I will only be
able to touch on a few highlights.

A crucial observation, due to Waelbroeck\ccite{Wael} and
't~Hooft,\ccite{Thooftb} is that in classical (2+1)-dimensional
gravity, a Regge calculus lattice picture is exact.  This is a
consequence of the fact that classical solutions are flat; in
contrast to the more familiar (3+1)-dimensional case, a
representation by flat simplices is thus a complete description.
This, in turn, means that extra edges may be added at will, or
conversely that only a few edge lengths are needed to completely
specify the geometry.

Waelbroeck uses this fact to express the lattice action in terms of
precisely the number of parameters needed to determine a geometric
structure.  For the $\IR\times T^2$ topology, these are just the four
holonomies $a$, $b$, $\lambda$, and $\mu$.  While the resulting quantum
theory has not yet been fully developed, it seems very likely that the
result will be equivalent to that of covariant canonical quantization.

An alternative lattice approach starts with the old observation
of Ponzano and Regge\ccite{Ponzano} that the Regge calculus action
in three dimensions can be approximated as a sum over angular
momenta of $6$-$j$ symbols.  The picture is one of attaching an
angular momentum $j$ to each edge of a tetrahedral lattice; a
tetrahedron's contribution to the action depends on the corresponding
$6$-$j$ symbol.  The result is invariant under subdivision, again
reflecting the exactness of the lattice approximation in three
dimensions.  By using the quantum group analog of angular momentum,
Turaev and Viro\ccite{Turaev} have recently showed how to evaluate
the sums over $j$; in an appropriate limit, the
result is equivalent to the Ponzano-Regge action on the one
hand,\ccite{Williams} and to the Euclideanized version of an ISO(2,1)
Chern-Simons theory on the other.\ccite{Turaev,Ooguri}  This lattice
approach thus provides a bridge between the first and second order
formalisms.  While the implications for quantum gravity have not
been fully developed, the connections to topological field theory
and three-manifold invariants provide interesting directions.

A very recent article by Rovelli\ccite{Rovb} ties the Turaev-Viro
approach to the loop representation.  Rovelli argues that the natural
basis of lattice states discussed by Ooguri,\ccite{Oogb} based
on ``colorings'' or values of $j$ on the boundary of a three-manifold,
is identical to the loop representation basis of section 6, and that
values of $j$ are simply edge lengths in the loop representation.
This might cause some concern, since we saw above that the loop and
connection representations are inequivalent.  But the Turaev-Viro
lattice picture probably corresponds to {\em Euclideanized}
(2+1)-dimensional gravity, for which the loop transform (\ref{loopt})
is a Fourier transformation rather than a Laplace transform, making
the two representations equivalent.  Perhaps the lesson here is that
one must be very careful about analytically continuing
to Riemannian metrics.

\section{Conclusion and Prospects}

In ordinary quantum field theory, we are used to the idea that many
different approaches to quantization lead to the same results.
In quantum
gravity, as we have seen, this is no longer the case.  There seem to
be at least four inequivalent quantum theories of (2+1)-dimensional
gravity---the reduced phase space ADM theory, the theory arising
from Chern-Simons
or covariant canonical quantization, the loop representation, and the
Wheeler-DeWitt formalism---and more undoubtedly await discovery.  The
basic problem, I believe, is that the equivalence proofs in quantum
field theory rely on renormalizability and locality.  General
relativity is not renormalizable in 3+1 dimensions, and more to
the point, the fundamental physical variables are not local.  As
a result, a candidate for a quantum theory of gravity is likely
to depend sensitively on choices
of variables and techniques of quantization.

In one sense, this is discouraging news.  For now, we have no criteria
for picking out the ``right'' quantization, and it is frustrating
to realize that we may be working on formulations that are ultimately
irrelevant to the real universe.  On the other hand, this variability
is also a cause for hope, since it means that a failure of one
approach need not doom others.  And some general results---the
importance of the mapping class group, the possibility of describing
geometry concretely in terms of nonlocal observables, the key role
of the space of classical solutions---seem
not to depend on a particular choice of quantization.  It is
clear, however, that we need to find better physical criteria to
determine which approaches are most likely to be correct.

In this talk, I have only touched on a few of the avenues that have
been used to explore (2+1)-dimensional quantum gravity.  For instance,
I have not mentioned the work of Nelson and Regge on the algebra of
observables,\ccite{Nelson,Nelsonb} or the many uses of path integral
methods.\ccite{Witb,Hosoya,Carlipentropy}  In restricting myself to
one simple topology, I have also ignored the interesting issues of
topology change.\ccite{Witb,Hor,Amano}  But I hope this talk has
provided a starting point for what can be a truly fascinating
subject.

\vspace{.2cm}
\begin{flushleft}
\large\bf Acknowledgements
\end{flushleft}
This work was supported in part by U.S.\ Department of Energy grant
DE-FG03-91ER40674 and National Science Foundation grant PHY89-04035.


\begin{thebibliography}{99}
\bibitem{DJtH} S.~Deser, R.~Jackiw, and G.~'t~Hooft,
 \Ann{152} (1984) 220.
\bibitem{tHooft} G.~'t~Hooft, \CMP{117} (1988) 685.
\bibitem{Martinec} E.~Martinec, \PRD{30} (1984) 1198.
\bibitem{Wita} E.~Witten, \NPB{311} (1988) 46.
\bibitem{Witb} E.~Witten, \NPB{323} (1989) 113.
\bibitem{Carlip1} S.~Carlip, \NPB{324} (1989) 106.
\bibitem{Thurston} W.P.~Thurston, {\sl The Geometry and Topology of
 Three-Manifolds}, Princeton lecture notes (1979).
\bibitem{Ep} R.~D.~Canary, D.~B.~A.~Epstein, and P.~Green,
 in {\sl Analytical and Geometric Aspects of Hyperbolic Space},
 London Math.~Soc.~Lecture Notes Series {\bf 111}, ed.\ D.~B.~Epstein
 (Cambridge University Press, Cambridge, 1987).
\bibitem{Mess} G.~Mess, ``Lorentz Spacetimes of Constant Curvature,''
 Institut des Hautes Etudes Scientifiques preprint IHES/M/90/28 (1990).
\bibitem{Carlipmeas} S.~Carlip, \CQG{8} (1991) 5.
\bibitem{Goldman} W.~M.\ Goldman, {\sl Inv.\ Math.} {\bf 93} (1988) 557.
\bibitem{Abikoff} W.~Abikoff, {\sl The Real Analytic Theory of
 Teichm\"uller Space}, Lecture Notes in Mathematics {\bf 820}
 (Springer-Verlag, Berlin, 1980).
\bibitem{Carlipobs} S.\ Carlip, \PRD{42} (1990) 2647.
\bibitem{Unruh} W.~G.\ Unruh and P.\ Newbury, ``Solution to 2+1 Gravity
 in the Dreibein Formalism,'' University of British Columbia preprint
 (1993).
\bibitem{Salomonson} P.\ Salomonson, B.~S.\ Skagerstam, and A.\ Stern,
 \NPB{347} (1990) 769.
\bibitem{Moncrief} V.\ Moncrief, \JMP{30} (1989) 2907.
\bibitem{HosNak} A.~Hosoya and K.~Nakao, \CQG{7} (1990) 163.
\bibitem{York} J.~W.~York, \PRL{28} (1972) 1082.
\bibitem{HosNak2} A.\ Hosoya and K.\ Nakao, {\sl Prog.\ Theor.\ Phys.}
 {\bf 84} (1990) 739.
\bibitem{Maassforms} H.~Iwaniec, in {\it Modular Forms}, ed.\
 R.~A.~Rankin (Ellis Horwood Ltd., Chichester, 1984).
\bibitem{Kuchar} K.\ Kucha{\v r}, in {\sl Proc.\ of the 4th Canadian
 Conf.\ on General Relativity and Relativistic Astrophysics}, ed.\ G.\
 Kunstatter et al.\ (World Scientific, Singapore, 1992).
\bibitem{Ash1} A.~Ashtekar, {\sl Lectures on Nonperturbative Quantum
 Gravity} (World Scientific, Singapore, 1991).
\bibitem{Ash2} A.\ Ashtekar et al., \CQG{6} (1989) L185.
\bibitem{Carlipdirac} S.\ Carlip, \PRD{45} (1992) 3584.
\bibitem{Carlipord} S.\ Carlip, ``The Modular Group, Operator Ordering,
 and Time in (2+1)-dimensional Gravity,'' Davis preprint UCD-92-23
 (1992).
\bibitem{Rovelli} C.\ Rovelli, \PRD{42} (1990) 2638; \PRD{43} (1991) 442.
\bibitem{Ash3} A.\ Ashtekar and A.\ Magnon, {\sl Proc.\ Roy.\ Soc.\
 (London)} {\bf A346} (1975) 375.
\bibitem{CrnWit} C.\ Crnkovic and E.\ Witten, in {\sl Three Hundred
 Years of Gravity}, ed.\ S.~W.\ Hawking and W.\ Israel (Cambridge
 University Press, Cambridge, 1987.)
\bibitem{Smolin} C.~Rovelli and L.~Smolin, \NPB{331} (1990) 80.
\bibitem{Marolf} D.\ M.\ Marolf, ``Loop Representations for 2+1 Gravity
 on a Torus,'' Syracuse preprint SU-GP-93/3-1 (1993).
\bibitem{diBart} C.\ Di~Bartolo, R.\ Gambini, and J.\ Griego, ``The
 Extended Loop Group,'' Simon Bolivar preprint 92-0333 (1992).
\bibitem{DeWittb} B.~S.\ DeWitt, {\sl Phys.\ Rev.} {\bf 160} (1967)
 1113.
\bibitem{Woodard} R.~P.\ Woodard, \CQG{10} (1993) 483.
\bibitem{CarlipWdW} S.\ Carlip, work in progress.
\bibitem{Wael} H.~Waelbroeck, \CQG{7} (1990) 751; \PRL{64} (1990) 2222.
\bibitem{Thooftb} G.\ 't~Hooft, \CQG{9} (1992) 1335.
\bibitem{Ponzano} G.~Ponzano and T.~Regge, in {\em Spectroscopic and
 Group Theoretical Methods in Physics}, ed.\ F.~Block (North-Holland,
 Amsterdam, 1968).
\bibitem{Turaev} V.~G.\ Turaev and O.~Y.\ Viro, {\em Topology} {\bf 31}
 (1992) 865.
\bibitem{Williams} F.\ Archer and R.~M.\ Williams, \PLB{273} (1991)
 438.
\bibitem{Ooguri} H.\ Ooguri and N.\ Sasakura, \MPLA{6} (1991) 3591.
\bibitem{Rovb} C.\ Rovelli, ``The Basis of the
 Ponzano-Regge-Turaev-Viro-Ooguri Model is the Loop Representation
 Basis,'' Pittsburgh preprint pitt-gr-4-93 (1993).
\bibitem{Oogb} H.\ Ooguri, \NPB{382} (1992) 276.
\bibitem{Nelson} J.~E.\ Nelson, T.\ Regge, and F.\ Zertuche, \NPB{339}
 (1990) 516.
\bibitem{Nelsonb} J.~E.\ Nelson and T.\ Regge, \CMP{141} (1991) 211;
 \IJMPB{6} (1992) 2091; \CQG{9} (1992) S187.
\bibitem{Hosoya} Y.\ Fujiwara et al., \PRD{44} (1991) 1756, 1763.
\bibitem{Carlipentropy} S.~Carlip, \PRD{46} (1992) 4387; \CQG{10}
 (1993) 207.
\bibitem{Hor} G.~T.\ Horowitz, \CQG{8} (1991) 587.
\bibitem{Amano} K.\ Amano and S.\ Higuchi, \NPB{377} (1992) 218.
\end{thebibliography}
\end{document}